\documentclass[pss]{wiley2sp} 
\usepackage{amsmath}
\usepackage{bm}
\usepackage{braket}

\begin{document}

\title{Recursive Calculation of the Optical Response of
  Multicomponent Metamaterials}

\author{%
  W. Luis Moch\'an\textsuperscript{\Ast,\textsf{\bfseries 1}},
  Raksha Singla\textsuperscript{\textsf{\bfseries 1}},
  Lucila Ju\'arez\textsuperscript{\textsf{\bfseries 2}}, and
  Guillermo P. Ortiz\textsuperscript{\textsf{\bfseries 3}}
}

\mail{e-mail
  \textsf{mochan@fis.unam.mx}}

\institute{%
  \textsuperscript{1}\,Instituto de Ciencias F\'isicas,
  Universidad Nacional Aut\'onoma de M\'exico, Av. Universidad s/n,
  Col. Chamilpa, 62210 Cuernavaca, Morelos, M\'exico
  \\
  \textsuperscript{2}\,Department of Photonics, Centro de
  Investigaciones en \'Optica, Le\'on, Guanajuato, M\'exico
  \\
  \textsuperscript{3}\, Departamento de F{\'\i}sica, Facultad de
  Ciencias Exactas Naturales y Agrimensura,  Universidad Nacional del
  Nordeste, Corrientes, Argentina
}

\keywords{metamaterials, homogenization, dielectric function}

\abstract{\bf%

  We develop a recursive computational procedure to
  efficiently calculate the macroscopic dielectric function of
  multi-component metamaterials of arbitrary geometry and composition
  within the long wavelength approximation. Although the microscopic
  response of the system might correspond to non-Hermitian operators,
  we develop a
  representation of the microscopic fields and of the response, and we
  introduce an appropriate metric that makes all operators
  symmetric. This allows us to use a modified Haydock recursion,
  introducing complex Haydock coefficients that allow an efficient
  computation of the macroscopic response and the microscopic
  fields. We test our procedure comparing our results to
  analytical ones in simple systems, and verifying they obey a
  generalized multicomponent Keller' theorem and the Mortola and
  Stefé's theorem for four component metalic and dielectric systems.
 }

\maketitle   

\section{Introduction}\label{intro}

Metamaterials made up of a repeated pattern of one or more ordinary
materials within a host have optical properties that might differ
substantially from those of its components
\cite{shalaev_optical_2007}. According to their
geometry and composition they might display both electric and magnetic
resonances of dipolar and quadrupolar nature
\cite{cho_contribution_2008} around which their macroscopic
permittivity and permeability
may become negative, yielding an exotic negative refraction
\cite{shelby2001experimental,smith2004metamaterials,aydin2005investigation,hoffman2007negative,peng2007experimental}.
A usual typical geometry of these left-handed and other exotic
metamaterials is that of pairs of wires and split conducting rings within a
dielectric. Nevertheless, to
avoid the dissipation inherent within the conducting phases,
all-dielectric structures employing high index of refraction
have also been investigated
\cite{jahani_all-dielectric_2016,kivshar_all-dielectric_2018}, and it
has been shown that their
Mie like resonances may also be employed for guiding light and for
enhancing non-linear optical
effects. All
dielectric structures of appropriate shapes may also exhibit negative
dispersion \cite{vynck_all-dielectric_2009}.

The microscopic field within metamaterials may have small regions
where the field is very high. Small modifications of the composition at
these regions may produce notable macroscopic effects. Thus,
metamaterials have been used to develop different kinds of sensors for
different spectral regions
\cite{ming_huang_microwave_2011,la_spada_metamaterial-based_2011,chen_metamaterials_2012,wongkasem_chiral_2017}.
Many other known and emerging applications of metamaterials have been
reviewed recently \cite{pendry_photonics:_2006,baev_metaphotonics:_2015}.

The permeability and permittivity, as well as chiral properties of a
metamaterial may be obtained from its reflection and transmission
properties \cite{smith_determination_2002} and from the dispersion of
guided modes within
metamaterial waveguides \cite{chen_experimental_2006}, but more fundamentally,
from the frequency and wavevector dependence of an appropriately
defined spatially-dispersive macroscopic dielectric function
\cite{agranovich_linear_2004,silveirinha_metamaterial_2007,agranovich_electrodynamics_2009,alu_restoring_2011,alu_first-principles_2011,konovalenko_nonlocal_2019}.

A very efficient scheme for the calculation of the optical properties of
metamaterials has been developed for binary metamaterials by
exploiting an analogy between the macroscopic dielectric tensor and
the projected Green's function corresponding to a Hermitian
Hamiltonian \cite{mochan_efficient_2010} which may be obtained through
Haydock's recursive procedure
\cite{haydock_electronic_1972,haydock_recursive_1980,pettifor_recursion_1984}.
The method has been used to
study extraordinary transmission through perforated metallic
slabs \cite{mochan_efficient_2010}, to calculate plasmonic properties of odd
shaped metallic inclusions \cite{cortes_optical_2010}, to study
enhanced birrefrigency and dichroism in anisotropic metamaterials
\cite{mendoza_birefringent_2012}, for
the design and optimization of optical devices to control the
absorption \cite{ortiz_effective_2014} and polarization of light
\cite{mendoza_tailored_2016}, and to optimize electrical and optical
properties of semitransparent contacts \cite{toranzos_optical_2017}.
The method has also been generalized to account for retardation, yielding a
non-local macroscopic response from which the complete band
structure of photonic crystals may be obtained
\cite{perez-huerta_macroscopic_2013} and from which magnetic properties may
be extracted \cite{juarez-reyes_magnetic_2018}.

There has also surged interest in the nonlinear properties of
metamaterials \cite{lapine_colloquium:_2014,czaplicki} and
metasurfaces \cite{czaplicki}. The Haydock's recursive
approach has been extended to calculate the microscopic field and from
it the macroscopic non-linear optical response of metamaterials. In
particular, to obtain
the second harmonic generation spectra of metamaterials with
centrosymmetric components but noncentrosymmetric shapes
\cite{meza_second-harmonic_2019}.

Unfortunately, the efficient computational approach developed in
\cite{mochan_efficient_2010} is directly applicable only to binary
metamaterials, that is, to systems composed of exactly two different
materials $A$ and $B$. The reason for this limitation is that the
geometry of such systems may be decoupled from their composition and
described by a characteristic function $B(\bm r)$ whose value
is 1 for those points $\bm r$ that belong to region $B$, and 0 when
$\bm r$ does not, i.e., within region $A$. It is from this
characteristic function that a Hermitian operator is built,
regardless of the actual composition of $A$ and $B$, and of their
dielectric or conducting nature, their dispersion and
dissipation. The Haydock coefficients for this operator are readily
obtained and from them a closed expression for its macroscopic
response may be built. However, for multicomponent systems, one cannot
find such a Hermitian operator to describe the geometry, and the
dielectric response itself is not Hermitian in the presence of dissipation.
Given this limitation, many interesting systems seem to lie beyond the
possibilities of the recursive approach. For example, metasurfaces are
arrangements of patterned particles on a substrate \cite{yu_flat_2014}
that have been used to manipulate the refraction of light
\cite{yu_light_2011} and produce flat lenses
\cite{khorasaninejad_metalenses_2016}, compound
lenses \cite{khorasaninejad_metalenses_2016} and even fabricate
spin switchable holograms \cite{chen_spin-controlled_2018}. A
numerical study of metasurfaces would require at least three materials
corresponding to the particles, the substrate and the
ambient. Similarly, it has been shown that the field enhancement due to
resonant excitation of plasmonic particles may not decrease when
protected by a dielectric, if the dielectric presents a coexisting Mie
resonance \cite{liren_deng_enhancing_2019}. Arrays of metallic cores
coated by semiconductors may also display negative index of refraction
as an electric dipole plasmonic resonance might coexist with a
magnetic dipole Mie resonance
\cite{paniagua-dominguez_ultra_2013,paniagua-dominguez_metallo-dielectric_2011,abujetas_photonic_2015}.
The study of these coupled plasmonic-Mie resonances requires accounting
at least for a core, a coating and the ambient.

The purpose of the present paper is to generalize the efficient homogenization
procedure using Haydock's recursion, as presented in
Ref. \cite{mochan_efficient_2010}, in order to deal with periodic metamaterials
of arbitrary geometry and composition and with an arbitrary number of
components, or even with a dielectric response that varies
continuously in space. To that end, we realize that though the
dielectric response is not in general an Hermitian operator, it
corresponds to a symmetrical complex operator. Thus, we can cautiously
employ well know theorems of linear algebra provided we define an
appropriate Euclidean-like metric instead of the usual Hermitian
metric. However, this metric couples Bloch waves moving in opposite
directions, requiring us to introduce a spinor-like two-component
representation of the Bloch states, with one component for each of the opposing
propagation directions. Besides obtaining the macroscopic dielectric
response of the system, we can also calculate the microscopic electric
field, so our procedure can further be employed in non-linear
calculations. We restrict ourselves to the non-retarded,
long-wavelength approximation, though the same ideas can be applied to
fully retarded calculations.

In order to verify the suitability of our computational procedure, we
calculate the macroscopic response of various 2D multicomponent
systems and verify that our results are consistent with a generalized
\cite{ortiz_kellers_2018}  Keller's
theorem \cite{keller_theorem_1964}, and with
Mortola and Steffé's exact expression
\cite{s._mortola_two-dimensional_1985,milton_proof_2001}
for four-component chess-board systems.

The structure of the paper is the following. In Sec. \ref{theory} we
present our theory: In Subsec. \ref{ssmulti} we develop
our recursive approach to the calculation of the dielectric
function. In order to test our results, in Subsec. \ref{sscoated} we
obtain analytical approximate formulae for the response of a simple
multicomponent 
system, in Subsec. \ref{ssKeller} we present a generalized Keller's
theorem for multicomponent 2D systems and in Subsec. \ref{ssmortola}
we discuss a 2D four component system for which exact analytical
expressions are available. In Sec. \ref{results} we present numerical
results for a variety of 2D systems and verify that they agree with
analytical results in the appropriate limits, that in general they
obey Keller's theorem and that they are consistent with Mortola and Steffe's
expression. Finally, Sec. \ref{conclusions} is devoted to conclusions.

\section{Theory}\label{theory}
\subsection{Multicomponent metamaterials}\label{ssmulti}
In the non-retarded, long-wavelength limit, the longitudinal
projection of the macroscopic dielectric function of a periodic system
may be obtained
from \cite{mochan_electromagnetic_1985,mochan_electromagnetic_II_1985}
\begin{equation}\label{epsM}
  (\hat\epsilon_M^{LL})^{-1}=(\hat\epsilon^{LL})^{-1}_{aa},
\end{equation}
where the superscript $LL$ and the subscript $aa$ on an operator
$\hat{\mathcal O}$ denote the
application of longitudinal projectors $\hat{\mathcal
  P}^L$ and the application of spatial average projectors
$\hat{\mathcal P}_a$ on both sides of $\hat{\mathcal O}$. For a
periodic system with its fields
represented in reciprocal space we may express the
longitudinal and average projectors by the matrices
\begin{equation}\label{PL}
  \mathcal P^L_{\bm G\bm G'}=\hat{\bm G}_{\bm k}\hat{\bm G}_{\bm k}\delta_{\bm G\bm G'},
\end{equation}
and
\begin{equation}\label{Pa}
  \mathcal P^a_{\bm G\bm G'}=\delta_{\bm G\bm 0}\delta_{\bm G'\bm 0},
\end{equation}
where $\{\bm G\}$ is the reciprocal lattice and we abbreviate the unit vectors,
\begin{equation}\label{hatG}
  \hat{\bm G}_{\bm k}\equiv \frac{\bm k+\bm G}{|\bm k+\bm G|}
\end{equation}
where $\bm k$ is a small Bloch's wavevector which in the
long-wavelength approximation is assumed to be much smaller that $G$,
except for the case $\bm G=0$, for which we define the direction $\hat{\bm
  0}_{\bm k}\equiv\hat{\bm k}/k$. We remark that we may simplify our
calculations by reinterpreting the $LL$ {\em projection} in Eq. (\ref{epsM})
and similar equations below by the $LL$ {\em component}, representing
any operator $\hat{\mathcal O}^{LL}$ by the matrix $\hat
{\bm G}_{\bm k}\cdot\mathcal O_{\bm G\bm G'} \hat {\bm G}'_{\bm k}$.

For a system with only two components $A$ and $B$ we define a
characteristic function $B(\bm r)$ which takes the values 0 when $\bm
r\in A$ and 1 when $\bm r\in B$. In this case, we may write the {\em
  microscopic} dielectric function
\begin{equation}\label{epsvsu}
  \epsilon(\bm r)=\frac{\epsilon_A}{u}(u-B(\bm r)),
\end{equation}
where $\epsilon_\alpha$ is the dielectric function of component
$\alpha=A,B$ and
\begin{equation}\label{u}
  u=\frac{1}{1-\epsilon_B/\epsilon_A},
\end{equation}
is the {\em spectral variable}. From Eqs. (\ref{epsM}) and
(\ref{epsvsu}) it is clear that we only need the average projection of
the operator
\begin{equation}\label{Gu}
  \hat{\mathcal G}(u)=(u-\hat B^{LL})^{-1},
\end{equation}
which plays the role of a Green's function for the operator $\hat
B^{LL}$, the longitudinal projection of the charateristic function.
The spectral variable $u$ would then play the role of a complex {\em
  energy} which depends
on the dielectric functions of both media, which in turn are generally complex
valued functions of the frequency. As $\hat
B^{LL}$ is a Hermitian operator, it can be represented as a
tridiagonal real matrix with diagonal elements $a_n$, and subdiagonal
and supradiagonal elements $b_n$, its Haydock coefficients, in a basis
of Haydock states $\ket{n}$ obtained from an initial {\em macroscopic} state
$\ket{0}$ by repeatedly applying $\hat B^{LL}$ and orthonormalizing the
resulting state, i.e., defining
\begin{equation}\label{haydockstep}
  \hat B^{LL}\ket{n}\equiv b_{n+1}\ket{n+1}+a_n\ket{n}+b_{n}\ket{n-1},
\end{equation}
with the condition
\begin{equation}\label{orthonorm}
  \braket{n|m}=\delta_{nm}.
\end{equation}
The resulting response is given by the continued fraction
\begin{equation}\label{epsMH}
  \epsilon_M^{LL}= \frac{\epsilon_A}{u}\left(u-a_0 -
    \frac{b_1^2}{u-a_1-
      \frac{b_2^2}{u-a_2-\frac{b_3^2}{\ddots}}}\right).
\end{equation}
Details of this procedure may be seen in Ref. \cite{mochan_efficient_2010}.

For multicomponent metamaterials the procedure above does not work, as
the geometry of the system would no longer be described by a
single characteristic function, and if we introduce several characteristic
functions, one for each component, then it wouldn't be possible to
represent all of them by tridiagonal matrices in the same basis. One
way out of this difficulty is to use the longitudinal part of the
microscopic dielectric function $\hat\epsilon^{LL}$ as the operator to use
in Haydock's recursion. If we replace the recursion
(\ref{haydockstep}) by
\begin{equation}\label{haydockstep1}
  \hat \epsilon^{LL}\ket{n}\equiv
  b_{n+1}\ket{n+1}+a_n\ket{n}+b_{n}\ket{n-1},\quad \text{(ND)}
\end{equation}
then the macroscopic response would be given by
\begin{equation}\label{epsM1}
  \epsilon_M^{LL}= \left(a_0 -
    \frac{b_1^2}{a_1-
      \frac{b_2^2}{a_2-\frac{b_3^2}{\ddots}}}\right).\quad \text{(ND)}
\end{equation}
Nevertheless, this procedure would only work in the absence of
dissipation, when $\epsilon(\bm r)$ is real and $\hat\epsilon^{LL}$ is
a Hermitian operator. Otherwise, there would be no reason for
Eq. (\ref{haydockstep1}) 
to contain only three terms on its RHS with
real coefficients nor for
its first and third terms to contain coefficients from the same set
$\{b_n\}$. We would have instead
\begin{equation}\label{haydockstep2}
  \begin{split}
    \hat \epsilon^{LL}\ket{n}\equiv&
    b_{n+1}\ket{n+1}+a_n\ket{n}+c_{n}\ket{n-1}\\
    &+d_{n}\ket{n-2}+\ldots
  \end{split}
\end{equation}
with complex coefficients $a_n$, $b_n$, $c_n$, $d_n$\ldots, and
Eq. (\ref{epsM1}) would no longer hold. For this reason we flagged
Eq. (\ref{haydockstep1}) and (\ref{epsM1}) with ND (no
dissipation).

We notice that even when there is dissipation, the longitudinal
dielectric function is a symmetrical operator. To show this, we chose
an {\em Euclidean} scalar product between states
\begin{equation}\label{scalarr}
  \braket{\phi|\psi}\equiv \int d^3\bm r\, \phi(\bm r)\psi(\bm r)
\end{equation}
where $\phi(\bm r)$ and $\psi(\bm r)$ are the {\em wavefunctions} that
represent the {\em states} $\ket{\phi}$ and $\ket{\psi}$ in real
space. Notice that in Eq. (\ref{scalarr}) we didn't conjugate $\phi(\bm
r)$ as we would have done had we chosen a {\em Hermitian}
product. We can express this scalar product in reciprocal
space as
\begin{equation}\label{scalarq}
  \braket{\phi|\psi}\equiv \int \frac{d^3\bm q}{(2\pi)^3} \phi(-\bm
  q)\psi(\bm q)=\int \frac{d^3\bm q}{(2\pi)^3} \phi(\bm
  q)\psi(-\bm q),
\end{equation}
where we define the Fourier transform $\zeta(\bm q)$ of any function
$\zeta(\bm r)$ through
\begin{equation}\label{Fourier}
  \zeta(\bm r)\equiv \int \frac{d^3\bm q}{(2\pi)^3} \zeta(\bm q)e^{i\bm
    q\cdot\bm r}.
\end{equation}
Notice in Eq. (\ref{scalarq}) the minus sign in the argument of
$\phi$ instead of the its conjugate as in Parseval's theorem. Then, we
may compute a matrix element of $\hat\epsilon^{LL}$ 
as
\begin{equation}\label{phiepspsi}
  \braket{\phi|\hat\epsilon^{LL}|\psi}=-\int\frac{d^3\bm
    q}{(2\pi)^3}\int\frac{d^3\bm q'}{(2\pi)^3}\phi(\bm q)\hat{\bm
    q}\cdot \epsilon(-\bm q-\bm q')\hat{\bm q}\psi(\bm q),
\end{equation}
where $\epsilon(\bm q)$ is the Fourier transform of $\epsilon(\bm r)$.
Clearly,
$\braket{\phi|\hat\epsilon^{LL}|\psi}=\braket{\psi|\hat\epsilon^{LL}|\phi}$
showing that the operator $\hat\epsilon^{LL}$ is symmetric under the
appropriate scalar product.

Notice that for a periodic system, $\epsilon(\bm r)=\epsilon(\bm r+\bm
R)$ may be written as a Fourier {\em series} with coefficients
\begin{equation}\label{epsG}
  \epsilon_{\bm G}=\int_{\text{UC}} \frac{d^3r}{\Omega} \epsilon(\bm
  r) e^{-i\bm G\cdot\bm r},
\end{equation}
related to the Fourier transform 
$\epsilon(\bm
q)=(2\pi)^3 \sum_{\bm G} \epsilon_{\bm G} \delta(\bm q-\bm G)$, where $\{\bm R\}$
is the Bravais lattice, $\{\bm G\}$ its reciprocal lattice, and UC
indicates that the integral is over a unit cell, whose volume is
$\Omega$. Thus, we can write Eq. (\ref{phiepspsi}) as
\begin{equation}\label{phiepspsi1}
  \begin{split}
  \braket{\phi|\hat\epsilon^{LL}|\psi}=&\int_{\text{BZ}}\frac{d^3\bm
    k}{(2\pi)^3}\sum_{\bm G}\sum_{\bm G'} \phi(-\bm k-\bm G)\\
  &\hat{\bm
    G}\cdot \epsilon_{\bm G-\bm G'}\hat{\bm G'}\psi(\bm k+\bm G'),
  \end{split}
\end{equation}
where we replaced the wavevector $\bm q$ by the sum of a Bloch's
vector $\bm k$ and some reciprocal vector $\bm G$, and BZ indicates
that the integral is over the first Brillouin zone. Ordinarily, in a
periodic system, the normal modes of the system may be chosen as Bloch
waves with a single Bloch's vector $\bm k$. However, our chosen
metric couples $\bm k$ to $-\bm k$. Thus we will consider
simultaneously states with given Bloch's vectors $\pm\bm k$ and denote
them using a spinor-like notation as,
\begin{equation}\label{spinor}
  \ket{\zeta}\to\left(
  \begin{array}{c}
    \zeta(\bm k+\bm G)\\
    \zeta(-\bm k+\bm G)
  \end{array}
  \right).
\end{equation}
Consequently, we represent the dielectric response as a $2\times2$
matrix,
\begin{equation}\label{epsspin}
  \hat\epsilon^{LL}\to\left(
  \begin{array}{cc}
    \hat{\bm G}_{\bm k}\cdot\epsilon_{\bm G-\bm G'}\hat{\bm G'}_{\bm
      k}&0\\
    0&\hat{\bm G}_{-\bm k}\cdot\epsilon_{\bm G-\bm G'}\hat{\bm G'}_{-\bm k}
  \end{array}
  \right).
\end{equation}
The scalar product (\ref{scalarq}) becomes
\begin{equation}\label{scalarspin}
  \begin{split}
    \braket{\phi|\psi}=\sum_G&\left(\phi(-\bm k-\bm G)\psi(\bm k
    +\bm G)
    \right.
    \\
    &
    +\left.\phi(\bm k-\bm G)\psi(-\bm k+\bm G)\right).
  \end{split}
\end{equation}
From Eqs. (\ref{spinor})--(\ref{scalarspin}) we get
\begin{equation}\label{phiepspsis}
  \begin{split}
  \braket{\phi|\hat\epsilon^{LL}|\psi}&=
  \sum_{\bm G\bm G'}\left(\phi(-\bm
  k-\bm G) \hat{\bm G}_{\bm k}\cdot\epsilon_{\bm G-\bm G'}\hat{\bm
    G'}_{\bm k}\psi(\bm k+\bm G')\right.
  \\
  &+ \left.
  \phi(\bm k-\bm G) \hat{\bm G}_{-\bm k}\cdot\epsilon_{\bm G-\bm G'}\hat{\bm
    G'}_{-\bm k} \psi(-\bm k+\bm G')\right).
  \end{split}
\end{equation}

Using Eqs. (\ref{spinor})--(\ref{phiepspsis}) we can proceed to build
a Haydock's representation of the operator $\hat\epsilon^{LL}$. We
start from a couple of macroscopic states representing
longitudinal waves propagating in the directions $\pm\hat{\bm k}$,
corresponding to the starting spinor
\begin{equation}\label{zero}
  \ket{0}\to\frac{1}{\sqrt2}\left(
  \begin{array}{c}
    1\\1
  \end{array}
  \right)\delta_{\bm G\bm 0},
\end{equation}
normalized according to Eq. (\ref{scalarspin}). We also define a state
$\ket{-1}\to 0$. Then we repeatedly apply $\hat\epsilon^{LL}$ using
the matrix representation (\ref{epsspin}) and we orthonormalize the
resulting states to the previously obtained states, through Haydock's
recursion
\begin{equation}\label{haydockstep3}
    b_{n+1}\ket{n+1}=\hat
    \epsilon^{LL}\ket{n}-a_n\ket{n}-b_{n}\ket{n-1},
\end{equation}
where we demand
\begin{equation}\label{orthonorm1}
  \braket{n|m}=\delta_{nm}
\end{equation}
using the product (\ref{scalarspin}).
Thus,
\begin{equation}\label{an}
  a_n=\braket{n|\hat\epsilon^{LL}|n}
\end{equation}
and
\begin{equation}\label{bnp1}
  \begin{split}
    b_{n+1}^2=&(\bra{n}\hat\epsilon^{LL}-a_n\bra{n}-b_{n}\bra{n-1})
    \\
    &(\hat\epsilon^{LL}\ket{n}-a_n\ket{n}-b_{n}\ket{n-1}).
  \end{split}
\end{equation}
We remark that the symmetry of $\hat\epsilon^{LL}$ guarantees that the
coefficient of $\ket{n-1}$ is $b_n$, that there are no more terms in
Eq. (\ref{haydockstep3}) and that the resulting state $\ket{n+1}$ is
implicitly orthogonal to all previous states $\ket{0}\ldots\ket{n-1}$
even though we only orthogonalize it explicitly to $\ket{n}$, except
for the accumulation of numerical errors, which would have to be
handled in the implementation
\cite{simon_analysis_1984,horst_d._simon_lanczos_1984,z._bai_lanczos_2000}.
In analogy to Ref. \cite{mochan_efficient_2010}, the products by
$\hat{\bm G}_{\pm\bm k}$ and $\hat{\bm G}'_{\pm\bm k}$ in
Eq. (\ref{phiepspsis}) may be
performed in reciprocal space, while the convolution with
$\epsilon_{\bm G-\bm G'}$ may be replaced by a simple multiplication with
$\epsilon(\bm r)$ in real space, so that we may apply the operator
$\hat\epsilon^{LL}$ without involving any large matrix product.
The Haydock coefficients in Eqs. (\ref{an}) and (\ref{bnp1})
are not guaranteed to be real and positive as those in
Eq. (\ref{haydockstep}) and may be complex valued. As in
Ref. \cite{mochan_efficient_2010}, in the orthonormal basis
$\{\ket{n}\}$ the microscopic longitudinal
dielectric function is represented by a tridiagonal symmetric matrix
\begin{equation}\label{tridiag}
  \hat\epsilon^{LL}\to T_{nn'}=\left(
  \begin{array}{ccccc}
    a_0&b_1&  0&  0&  \ldots\\
    b_1&a_1&b_2&  0&  \ldots\\
    0  &b_2&a_2&b_3&  \ldots\\
    \vdots&\vdots&\vdots&\vdots&\ddots
  \end{array}
  \right),
\end{equation}
from which Eq. (\ref{epsM}) allows to extract the macroscopic response
\begin{equation}\label{epsM2}
  \epsilon_M^{LL}= \left(a_0 -
    \frac{b_1^2}{a_1-
      \frac{b_2^2}{a_2-\frac{b_3^2}{\ddots}}}\right).
\end{equation}
Notice that Eq. (\ref{epsM2}) seems identical to (\ref{epsM1}), but
its Haydock coefficients are different, as they are obtained by using
spinor-like states and an Euclidean metric. Thus, Eq. (\ref{epsM2})
may be used for arbitrary compositions, including multiple disperssive
and dissipative media or even a continuosly varying complex response
$\epsilon(\bm r)$.

By identifying the longitudinal displacement field $\bm D^L$
with an external macroscopic field and thus with no spatial
fluctuations, we may represent it in Haydock's basis as a column
vector with components $\bm D^L\to d_n=D^L \delta_{n0}$. We may expand
the longitudinal electric field in the same basis as $\bm E^L\to e_n$,
and solve the tridiagonal system
\begin{equation}\label{TEvsD}
  \sum_{n'} T_{nn'} e_{n'}=d_{n'}
\end{equation}
for the unknowns $e_n$ to obtain a representation of the microscopic
electric field $\bm E^L\to\sum e_n\ket{n}$ which may be translated into
reciprocal or real space to obtain $\bm E^L(\bm k+\bm G)$ or $\bm
E^L(\bm r)$.

We have implemented the formalism above as a set of modules written in
the Perl programming language, using its
Perl Data Language (PDL) \cite{glazebrook97pdl} extension for efficient
numerical calculations, and the Moose \cite{moose} object system, and
we have incorporated them into the publicly available package {\em
  Photonic} \cite{photonic}.  

\subsection{Coated clylinders}\label{sscoated}
A simple system to test our approach above is that of a lattice
of coated cylindrical particles. Consider a single multilayered
cylindrical particle  with a core ($p=1$) covered by $N-1$ coaxial shells
($p=2\ldots N$) within vacuum ($p=N+1$). Each layer is characterized by an
outer radius $a_p$ and a dielectric function $\epsilon_p$. The system
is subject
to an external field
$\bm E_{\text{ex}}=E_{\text{ex}} \hat x$.
The potential within each layer may be written as
\begin{equation}\label{layer}
  \phi_p(\bm r)=(A_p r + B_p/r)\cos\theta
\end{equation}
in polar coordinates, where, using the symmetry of the system, we
restricted ourselves to the angular momentum $l=1$ of the external
potential. The boundary conditions at the $p$-th boundary 
may be written as
\begin{equation}\label{transfer}
  \left(
  \begin{array}{c}
    A_{p+1}\\B_{p+1}
  \end{array}
  \right) = \mathcal M_p \left(
  \begin{array}{c}
    A_{p}\\B_{p}
  \end{array}
  \right),
\end{equation}
where we introduced the {\em transfer matrix}
\begin{equation}\label{transferM}
  M_p=\frac{1}{2} \left(
  \begin{array}{cc}
    \frac{\epsilon_{p+1} + \epsilon_{p}} {\epsilon_{p+1}} &
    \frac{\epsilon_{p+1} - \epsilon_{p}} {a_p^2\epsilon_{p+1}}\\
    \frac{\epsilon_{p+1} - \epsilon_{p}} {\epsilon_{p+1}}a_p^2 &
        \frac{\epsilon_{p+1} + \epsilon_{p}} {\epsilon_{p+1}}
  \end{array}
  \right).
\end{equation}
Using Eq. (\ref{transfer}) repeatedly, we may relate
\begin{equation}\label{Np1}
  \left(
  \begin{array}{c}
    A_{N+1}\\B_{N+1}
  \end{array}
  \right) = \mathcal M \left(
  \begin{array}{c}
    A_{1}\\B_{1}
  \end{array}
  \right),
\end{equation}
with $\mathcal M=\mathcal M_N\mathcal M_{N-1}\ldots\mathcal
M_2\mathcal M_1$. 
As we may identify $A_{N+1}=-E_{\text{ex}}$ and $B_{N+1}=2p$, with $p$
the total dipole moment per unit length, and as $B_1=0$ to avoid a
singularity at $r=0$, from Eq. (\ref{Np1}) we may obtain the
polarizability per unit length of the particle
\begin{equation}\label{alpha}
  \alpha=\frac{p}{E_{\text{ex}}}=-\frac{\mathcal M_{21}}{2\mathcal M_{11}}.
\end{equation}

For a square array of such coated cylinders we may approximate the
macroscopic dielectric response through the Claussius-Mossotti
2D relation
\begin{equation}\label{epsM3}
  \epsilon_M=\frac{1+2\pi n\alpha}{1-2\pi n\alpha}
\end{equation}
where $n$ is the number density. We expect this expression to hold as
long as the distance between cylinders is not so short as to allow
exciting multipoles higher than the dipole.

\subsection{Keller's theorem}\label{ssKeller}

In order to further test our result (\ref{epsM2}) we will show below that they
satisfy a generalization of Keller's theorem \cite{ortiz_kellers_2018}
for multicomponent metamaterials, which we prove in a simple (limited)
form below. Consider a 2D metamaterial with three or more components $A$,
$B$, $C$\ldots, each characterized by a dielectric function
$\epsilon_A$, $\epsilon_B$, $\epsilon_C$\ldots Then, we write its dielectric
function as
\begin{equation}\label{epsABC}
  \epsilon\equiv\epsilon(\epsilon_A,\epsilon_B,\epsilon_C,\ldots) =
  \epsilon_A A + \epsilon_B B + \epsilon_C C + \ldots,
\end{equation}
where we introduced characteristic functions $A(\bm r)$, $B(\bm r)$,
$C(\bm r)$\ldots, that take the value 1 when $\bm r$ lies within the
corresponding region $A$, $B$, $C$\ldots and 0 otherwise. We expect
the use of the same
letters to denote materials, regions and characteristic functions will
not be confusing, as their use may be distinguished by context. In the
absence of external charge and neglecting retardation, the microscopic
displacement and electric fields are solenoidal
and irrotational respectively,
\begin{equation}\label{divD}
  \begin{array}{cc}
    \nabla\cdot \bm D=0,&\nabla\times\bm E=0,
  \end{array}
\end{equation}
and they are related through
\begin{equation}\label{DvsE}
  \bm D=\epsilon(\epsilon_A,\epsilon_B,\epsilon_C\ldots)\bm
  E=\epsilon\bm E.
\end{equation}

Now consider the rotated fields $\bm D^R(\bm r)\equiv \mathcal R \bm
D(\bm r)$ and $\bm E^R(\bm r)\equiv \mathcal R \bm E(\bm r)$, where
\begin{equation}\label{rotation}
  \mathcal R\equiv\left(
  \begin{array}{cc}
    0&1\\-1&0
  \end{array}\right)
\end{equation}
is a rotation matrix by $\pi/2$. Notice that we
rotate the fields but not their application point $\bm r$. As
$\mathcal R$ coincides with the Levi-Civita symbol in 2D, then
\begin{equation}\label{divER}
  \begin{array}{cc}
    \nabla\cdot \bm E^R=0,&\nabla\times\bm D^R=0,
  \end{array}
\end{equation}
which are similar to Eqs. (\ref{divD}) but with the substitutions
\begin{equation}\label{subst}
  \begin{array}{cc}
    \bm D\to\tilde {\bm D}\equiv\bm E^R,&\bm E\to\tilde{\bm
      E}\equiv \bm D^R.
  \end{array}
\end{equation}
Notice that
\begin{equation}\label{ERvsDR}
  \tilde{\bm D} =
  \epsilon(1/\epsilon_A,1/\epsilon_B,1/\epsilon_C\ldots)\tilde{\bm
    E}\equiv \tilde\epsilon \tilde{\bm E}.
\end{equation}
Thus, $\tilde{\bm D}$ and $\tilde{\bm E}$ obey the same equations as
$\bm D$ and $\bm E$ but they are related through a microscopic
dielectric response $\tilde\epsilon$ identical to that in
Eq. (\ref{epsABC}) but for 
the replacements $\epsilon_A\to1/\epsilon_A$,
$\epsilon_B\to1/\epsilon_B$, $\epsilon_C\to1/\epsilon_C$\ldots

Through a homogenization procedure, such as using Eq. (\ref{epsM}),
from Eqs. (\ref{DvsE}) and (\ref{ERvsDR}) we obtain
$\bm D_M=\bm\epsilon_M\bm E_M$ and $\tilde{\bm D}_M = \tilde{\bm
  \epsilon}_M \tilde{\bm E}_M$,
where $\bm\epsilon_M$ is the macroscopic dielectric tensor of the original
system and $\tilde{\bm \epsilon}_M$ is the corresponding response of the
system obtained from the original one by replacing the response of each
component by its inverse. Then we may write
\begin{equation}\label{preK}
  \begin{split}
    \bm E_M=&\mathcal R^{-1}\tilde{\bm D}_M=
    \mathcal R^{-1}\tilde{\bm \epsilon}_M\tilde{\bm E}_M
    \\
    &=
    \mathcal R^{-1}\tilde{\bm \epsilon}_M\mathcal R{\bm D_M}=
    \mathcal R^{-1}\tilde{\bm \epsilon}_M\mathcal R\bm \epsilon_M{\bm E_M},
  \end{split}
\end{equation}
and cancelling $\bm E_M$ we finally obtain
\begin{equation}\label{Keller}
  \tilde{\bm\epsilon}_M^R\bm \epsilon_M=\bm 1,
\end{equation}
where $\tilde{\bm \epsilon}_M^R=\mathcal
R^{-1}\tilde{\bm\epsilon}_M\mathcal R$.
Thus the original macroscopic
response times the rotated macroscopic response of the system with the
reciprocal dielectric functions yields the unit tensor. This is
Keller's theorem for multicomponent metamaterials in 2D.

\subsection{Mortola and Steffe's theorem}\label{ssmortola}
Consider now a 2D system made of a square lattice whose unit cell is
divided into four identical squares that are occupied from left to
right, top to bottom, by four materials $A$, $B$, $C$, $D$. Mortola
and Steffe proposed an expression
\cite{s._mortola_two-dimensional_1985} for the macroscopic
conductivity of this system in terms of the conductivities of its
components. This expression was later proved by Milton
\cite{milton_proof_2001}. However, as argued in
\cite{ortiz_kellers_2018}, we expect that the correct expression for
finite frequencies is that written in terms of the dielectric
response, i.e.,
\begin{equation}\label{Mortola}
  \begin{split}
    \epsilon_M^{xx}=&\{[(\epsilon_A+\epsilon_C) (\epsilon_B+\epsilon_D)
      (\epsilon_A\epsilon_B\epsilon_C + \epsilon_B\epsilon_C\epsilon_D
      \\
      &+ \epsilon_C\epsilon_D\epsilon_A +
      \epsilon_D\epsilon_A\epsilon_B)]/[(\epsilon_A+\epsilon_B)
      (\epsilon_C+\epsilon_D)
      \\ &\times
      (\epsilon_A+\epsilon_B +\epsilon_C+\epsilon_D)]\}^{1/2}.
  \end{split}
\end{equation}
A similar expression holds for $\epsilon_M^{yy}$ obtained from
Eq. (\ref{Mortola}) by exchanging $B\leftrightarrow C$.

\section{Results}\label{results}

\begin{figure*}[htb]%
  \sidecaption
  \includegraphics*[width=.7\textwidth]{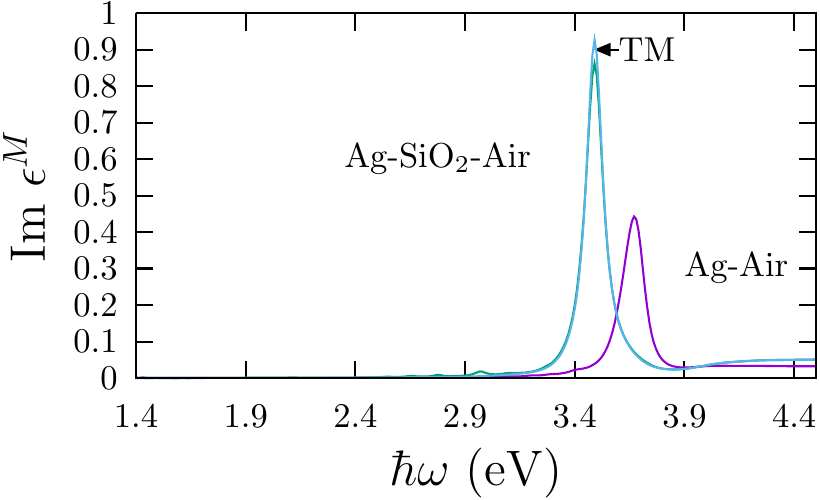}
    \caption{%
      Imaginary part of the macroscopic dielectric function
      $\epsilon_M$ of a square lattice of Ag cylinders of radius
      $a_{\text{Ag}}=0.1L$ in vacuum, where $L$ is the lattice parameter. One
      curve corresponds to uncoated cylinders and the other to
      cylinders coated by a SiO$_2$ shell with outer radius
      $a_{\text{SiO}_2}=0.25L$. We show results obtained
      analytically through Eq. (\ref{epsM3}) and the transfer matrix
      formalism (TM) and numerically through the procedure developed
      in Subsec. \ref{ssmulti}.
      \label{fcoated}}
\end{figure*}

In Figure \ref{fcoated} we show the imaginary part of the dielectric
function of a square lattice of thin coated and uncoated Ag cylinders
of radius 
$a_{\text{Ag}}=0.1L$ within vacuum, where $L$ is the lattice parameter. As
the cylinders are very thin, their mutual interaction is
negligible. Thus, in the case of the uncoated cylinders, there is a
peak around $\hbar\omega\approx 3.7$eV which corresponds to the
surface plasmon of an isolated Ag cylinder, given by $\epsilon_{\text
  {Ag}}=-1$. If the cylinder is coated by a SiO$_2$ layer of outer
radius $a_{\text{SiO}_2}=0.25L$ the peak is redshifted. The
analytical result based on the Claussius-Mossotti relation using the
polarizability given by Eq. (\ref{alpha}) based on a transfer matrix
formalism agrees quite closely with the numerical calculation based on
Haydock's recursion for the case of coated cylinders and is
indistinguishable for the case of uncoated cylinders. The numerical
calculation was done using a $401\times401$ grid and with up to 200
pairs of Haydock coefficients.

\begin{figure*}[htb]%
  \sidecaption
  \includegraphics*[width=.7\textwidth]{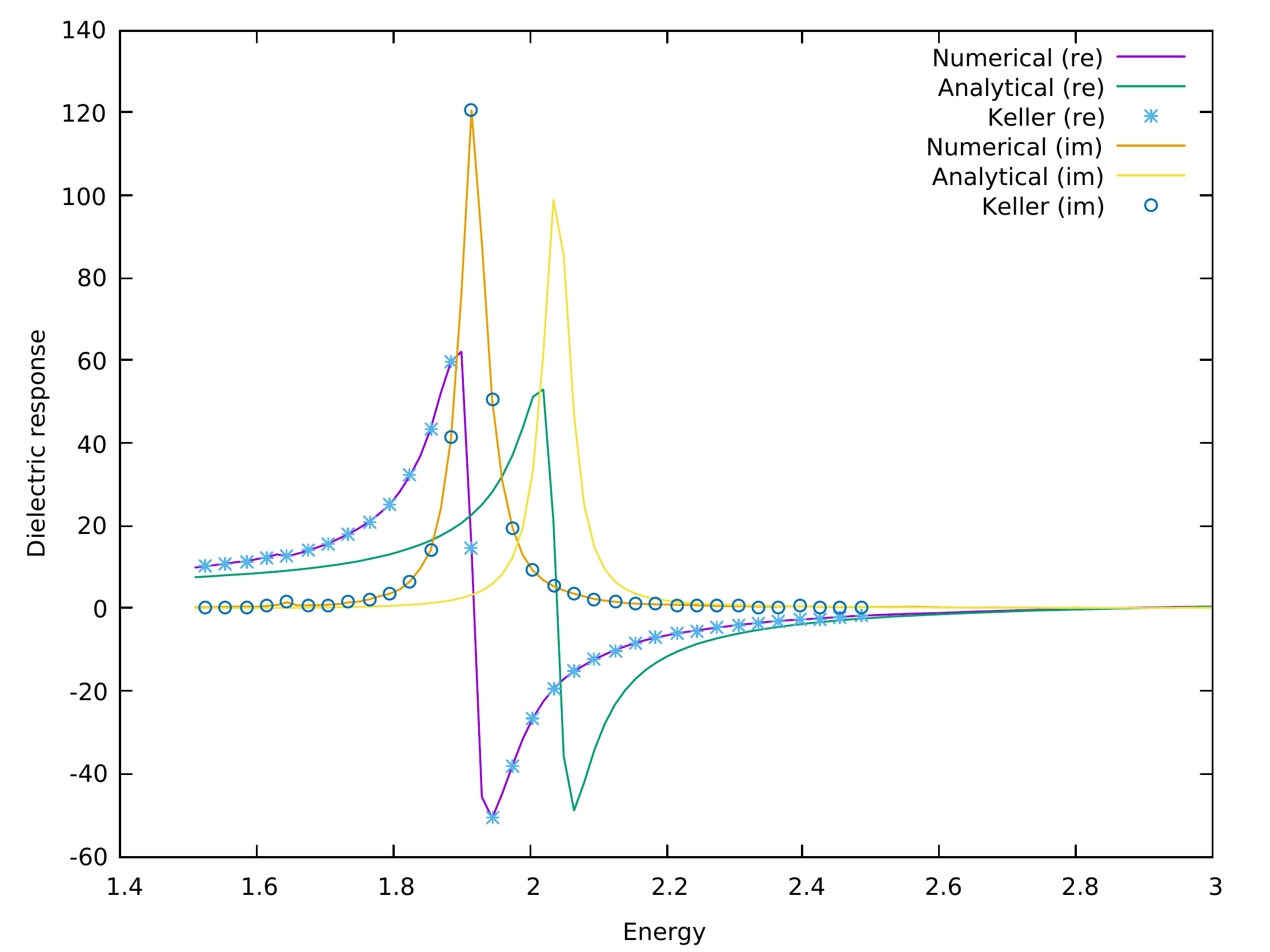}
    \caption{%
      Real and imaginary part of 
      the macroscopic dielectric function
      $\epsilon_M$ of a square lattice of SiO$_2$ cylinders of radius
      $a_{\text{SiO}_2}=0.3L$ covered by shells of Ag with outer radius
      $a_{\text{Ag}}=0.45L$, where $L$ is the lattice parameter. We
      show results obtained 
      analytically and numerically through the procedure developed
      in Subsec. \ref{ssmulti}. We also show results obtained
      numerically by applying Keller's theorem. 
    \label{fglassAg}}
\end{figure*}
Fig. \ref{fglassAg} shows the real and imaginary parts of the
macroscopic dielectric function for a system similar to that in
Fig. \ref{fcoated} but with an SiO$_2$ core of radius
$a_{\text{SiO}_2}=0.3L$ covered by an Ag shell of outer radius
$a_{\text{Ag}}=0.45L$. As neighboring cylinders are closer together
than in Fig. \ref{fcoated} dipolar and higher multipoles may couple
together. Thus, the extension (\ref{epsM3}) of the Claussius-Mossotti
formalism may not be accurate. The response obtained from the
numerical calculation has a peak around 1.92eV further red-shifted
from that of the isolated cylinder than the peak of the analytical
calculation around 2.04eV. Nevertheless, a numerical calculation based
on Keller's theorem, Eq. (\ref{Keller}), obtained by inverting the
dielectric functions of the components, calculating the corresponding
dielectric function using our recursive formalism and inverting the
result, seems to agree perfectly with the straightforward numerical
calculation. Thus, our recursive procedure agrees with Keller's
theorem even for large inclusions and strong interactions. 

\begin{figure*}[htb]%
  \sidecaption
  \includegraphics*[width=.7\textwidth]{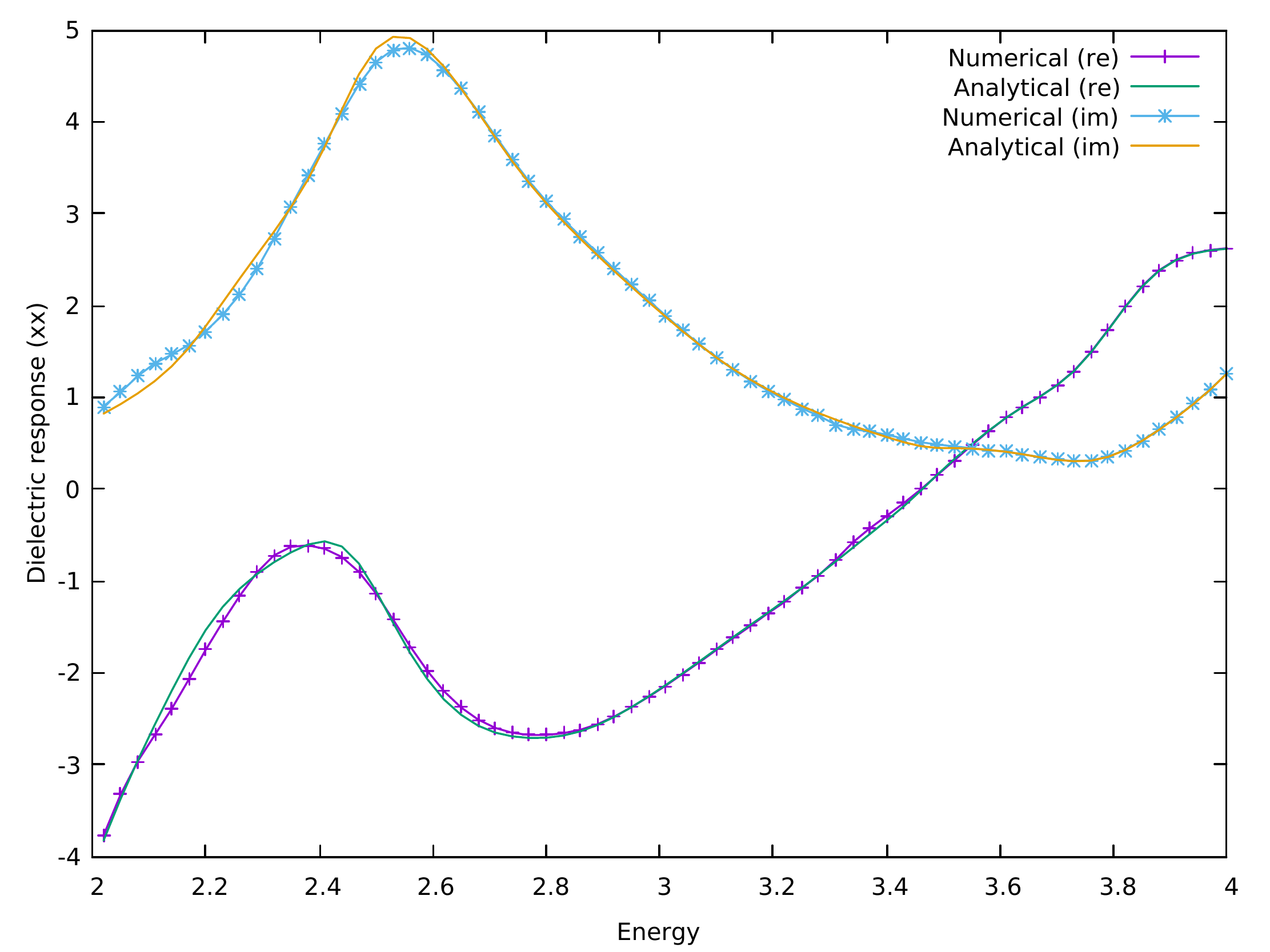}
    \caption{%
      Real and imaginary part of the component $\epsilon_M^{xx}$ of
      the macroscopic dielectric tensor of a four component
      metamaterial made of square prisms of Au, Ag, TiO$_2$ and
      SiO$_2$ arranged in a square unit cell repeated periodically
      with the geometry of a checkerboard, as illustrated by
      Fig. \ref{fcheckerboard}, as a function of photon energy
      $\hbar\omega$. The results obtained numerically using the
      procedure developed in Subsec. \ref{ssmulti} are compared to those
      obtained from the analytical formula presented in Subsec. \ref{ssmortola}
    \label{fmortola}}
\end{figure*}

\begin{figure}[htb]%
\includegraphics*[width=.4\textwidth]{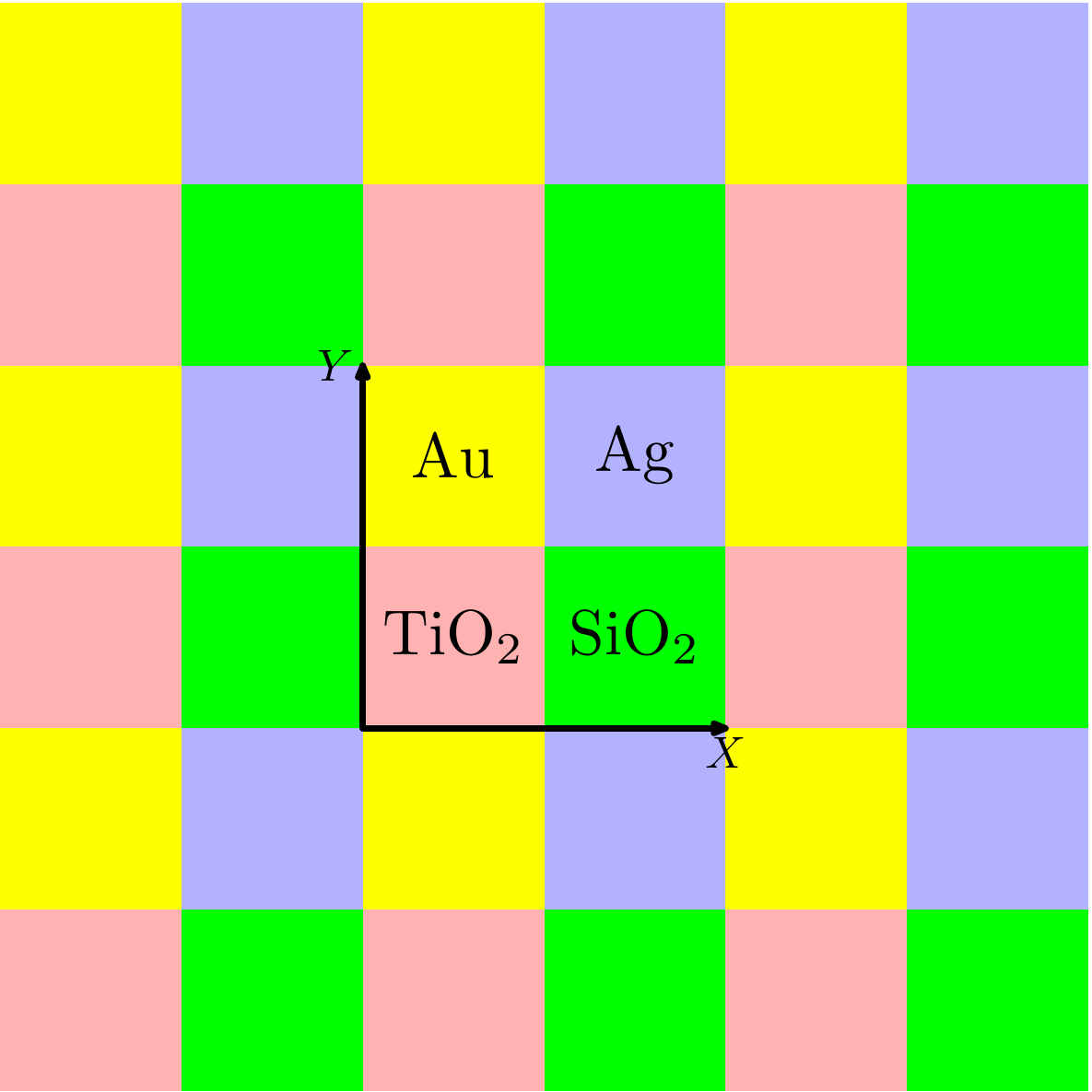}
\caption{%
  Geometry of the metamaterial corresponding to Fig. \ref{fmortola},
  consisting of the periodic repetition of a unit cell made up of four
  square prisms of Au, Ag, TiO$_2$ and SiO$_2$.}
\label{fcheckerboard}
\end{figure}

In Fig. \ref{fmortola} we show the real and imaginary parts of a
component $\epsilon_M^{xx}$  of the 
macroscopic dielectric tensor of a metamaterial made up of four
materials, Au, Ag, TiO$_2$ and SiO$_2$ filling square prisms occupying
a 2x2 block and repeated periódically in a checkerboard geometry, as
illustrated in Fig. \ref{fcheckerboard}. The 
calculation was performed using the procedure described in
Subsec. \ref{ssmulti} using a $201 \times 201$ grid and with up to 300
Haydock coefficient pairs. In the figure we also show the results of
an analytical calculation using the formula presented in Subsec. \ref{ssmortola}.
Notice the good agreement for both the real and imaginary parts for a
wide energy range.

\section{Conclusions}\label{conclusions}
We have developed a recursive procedure based on a Haydock's
representation that allows the efficient
calculation of the macroscopic dielectric function and the microscopic
fields of multicomponent metamaterials of arbitrary composition and
geometry. Our formalism admits materials that can be insulating, conducting,
transparent, opaque, dissipative, and/or dispersive. Although the
response of the system may be non-Hermitian, we could take advantage of its
symmetric nature by introducing an appropriate scalar product and
using a spinor-like representation of the fields. Though efficient,
the procedure developed here is not as fast as that for only two
materias, as in the current case the Haydock coefficients depend on
the composition and not only on the geometry. The results
presented here correspond to the non-retarded limit, though we
have verified that the same ideas may be extended to the retarded
region where they may even be applied to chiral systems. We have
prepared computational modules implementing our procedures and added
them to a publicly available software package. We tested our
formalism by calculating the response of simple 
systems for which approximate analytical formulae are available, and
by demonstrating that our results are consistent with some exact
conditions, namely, Keller's and Mortola and Steffe's theorems.

\begin{acknowledgement}
This work was supported by DGAPA-UNAM under grant IN111119. RS
acknowledges a scholarship from CONACyT.
\end{acknowledgement}

%
\bibliographystyle{pss}
\bibliography{biblio}

\providecommand{\WileyBibTextsc}{}
\let\textsc\WileyBibTextsc
\providecommand{\othercit}{}
\providecommand{\jr}[1]{#1}
\providecommand{\etal}{~et~al.}


\begin{thebibliography}{[10]}

\bibitem{shalaev_optical_2007}
 \textsc{V.\,M. Shalaev},
 \jr{Nat Photon} \textbf{1}(1), 41--48 (2007).


\bibitem{cho_contribution_2008}
 \textsc{D.\,J. Cho},  \textsc{F.~Wang},  \textsc{X.~Zhang},  and
  \textsc{Y.\,R. Shen},
 \jr{Phys. Rev. B} \textbf{78}(12), 121101 (2008).


\bibitem{shelby2001experimental}
 \textsc{R.\,A. Shelby},  \textsc{D.\,R. Smith},  and
  \textsc{S.~Schultz},
 \jr{Science} \textbf{292}(5514), 77--79 (2001).


\bibitem{smith2004metamaterials}
 \textsc{D.\,R. Smith},  \textsc{J.\,B. Pendry},  and  \textsc{M.\,C.
  Wiltshire},
 \jr{Science} \textbf{305}(5685), 788--792 (2004).


\bibitem{aydin2005investigation}
 \textsc{K.~Aydin},  \textsc{I.~Bulu},  \textsc{K.~Guven},
  \textsc{M.~Kafesaki},  \textsc{C.\,M. Soukoulis},  and
  \textsc{E.~Ozbay},
 \jr{New Journal of Physics} \textbf{7}(1), 168 (2005).


\bibitem{hoffman2007negative}
 \textsc{A.\,J. Hoffman},  \textsc{L.~Alekseyev},  \textsc{S.\,S. Howard},
  \textsc{K.\,J. Franz},  \textsc{D.~Wasserman},  \textsc{V.\,A. Podolskiy},
  \textsc{E.\,E. Narimanov},  \textsc{D.\,L. Sivco},  and
  \textsc{C.~Gmachl},
 \jr{Nature materials} \textbf{6}(12), 946 (2007).


\bibitem{peng2007experimental}
 \textsc{L.~Peng},  \textsc{L.~Ran},  \textsc{H.~Chen},  \textsc{H.~Zhang},
  \textsc{J.\,A. Kong},  and  \textsc{T.\,M. Grzegorczyk},
 \jr{Physical review letters} \textbf{98}(15), 157403 (2007).


\bibitem{jahani_all-dielectric_2016}
 \textsc{S.~Jahani} and  \textsc{Z.~Jacob},
 \jr{Nature Nanotechnology} \textbf{11}(1), nnano.2015.304 (2016).


\bibitem{kivshar_all-dielectric_2018}
 \textsc{Y.~Kivshar},
 \jr{Natl Sci Rev} \textbf{5}(2), 144--158 (2018).


\bibitem{vynck_all-dielectric_2009}
 \textsc{K.~Vynck},  \textsc{D.~Felbacq},  \textsc{E.~Centeno},  \textsc{A.\,I.
  Căbuz},  \textsc{D.~Cassagne},  and  \textsc{B.~Guizal},
 \jr{Phys. Rev. Lett.} \textbf{102}(13), 133901 (2009).


\othercit
\bibitem{ming_huang_microwave_2011}
 \textsc{{Ming Huang}} and  \textsc{{Jingjing Yang}},
Microwave {Sensor} {Using} {Metamaterials},
 in: Wave {Propagation},  (InTech, March 2011),
Dr. Andrei Petrin (Ed.).


\bibitem{la_spada_metamaterial-based_2011}
 \textsc{L.~La~Spada},  \textsc{F.~Bilotti},  and  \textsc{L.~Vegni},
 \jr{Progress In Electromagnetics Research B} \textbf{34}(January) (2011).


\bibitem{chen_metamaterials_2012}
 \textsc{T.~Chen},  \textsc{S.~Li},  and  \textsc{H.~Sun},
 \jr{Sensors} \textbf{12}(3), 2742--2765 (2012).


\bibitem{wongkasem_chiral_2017}
 \textsc{N.~Wongkasem},  \textsc{A.~Sonsilphong},  and
  \textsc{M.~Gonzalez},
 \jr{1} \textbf{44}(3), 178--181 (2017).


\bibitem{pendry_photonics:_2006}
 \textsc{J.~Pendry},
 \jr{Nat Mater} \textbf{5}(8), 599--600 (2006).


\bibitem{baev_metaphotonics:_2015}
 \textsc{A.~Baev},  \textsc{P.\,N. Prasad},  \textsc{H.~Ågren},
  \textsc{M.~Samoć},  and  \textsc{M.~Wegener},
 \jr{Physics Reports} \textbf{594}(September), 1--60 (2015).


\bibitem{smith_determination_2002}
 \textsc{D.\,R. Smith},  \textsc{S.~Schultz},  \textsc{P.~Markoš},  and
  \textsc{C.\,M. Soukoulis},
 \jr{Phys. Rev. B} \textbf{65}(19), 195104 (2002).


\bibitem{chen_experimental_2006}
 \textsc{H.~Chen},  \textsc{J.~Zhang},  \textsc{Y.~Bai},  \textsc{Y.~Luo},
  \textsc{L.~Ran},  \textsc{Q.~Jiang},  and  \textsc{J.\,A. Kong},
 \jr{Opt. Express, OE} \textbf{14}(26), 12944--12949 (2006).


\bibitem{agranovich_linear_2004}
 \textsc{V.\,M. Agranovich},  \textsc{Y.\,R. Shen},  \textsc{R.\,H. Baughman},
  and  \textsc{A.\,A. Zakhidov},
 \jr{Phys. Rev. B} \textbf{69}(16), 165112 (2004).


\bibitem{silveirinha_metamaterial_2007}
 \textsc{M.\,G. Silveirinha},
 \jr{Phys. Rev. B} \textbf{75}(11), 115104 (2007).


\bibitem{agranovich_electrodynamics_2009}
 \textsc{V.\,M. Agranovich} and  \textsc{Y.\,N. Gartstein},
 \jr{Metamaterials} \textbf{3}(1), 1--9 (2009).


\bibitem{alu_restoring_2011}
 \textsc{A.~Alù},
 \jr{Phys. Rev. B} \textbf{83}(8), 081102 (2011).


\bibitem{alu_first-principles_2011}
 \textsc{A.~Alù},
 \jr{Phys. Rev. B} \textbf{84}(7), 075153 (2011).


\bibitem{konovalenko_nonlocal_2019}
 \textsc{A.~Konovalenko},  \textsc{J.\,A.\,R. Avendaño},  \textsc{A.\,M.
  Blas},  \textsc{F.~Cervera},  \textsc{E.~Myslivets},  \textsc{S.~Radic},
  \textsc{J.\,S.\,D. Moreno},  and  \textsc{F.~Perez-Rodriguez},
 \jr{J. Opt.} (2019).


\bibitem{mochan_efficient_2010}
 \textsc{W.\,L. Mochán},  \textsc{G.\,P. Ortiz},  and  \textsc{B.\,S.
  Mendoza},
 \jr{Opt. Express, OE} \textbf{18}(21), 22119--22127 (2010).


\bibitem{haydock_electronic_1972}
 \textsc{R.~Haydock},  \textsc{V.~Heine},  and  \textsc{M.\,J. Kelly},
 \jr{J. Phys. C: Solid State Phys.} \textbf{5}(20), 2845 (1972).


\bibitem{haydock_recursive_1980}
 \textsc{R.~Haydock},
 \jr{Solid State Physics} \textbf{35}, 215 (1980).


\othercit
\bibitem{pettifor_recursion_1984}
 \textsc{D.\,G. Pettifor} and  \textsc{D.\,L. Weaire} (eds.),
The {Recursion} {Method} and {Its} {Applications}, Springer {Series} in {Solid}
  {State} {Sciences},  Vol.\,58 (Springer, Berlin, 1984).


\bibitem{cortes_optical_2010}
 \textsc{E.~Cortes},  \textsc{L.~Mochán},  \textsc{B.\,S. Mendoza},  and
  \textsc{G.\,P. Ortiz},
 \jr{phys. stat. sol. (b)} \textbf{247}(8), 2102--2107 (2010).


\bibitem{mendoza_birefringent_2012}
 \textsc{B.\,S. Mendoza} and  \textsc{W.\,L. Mochán},
 \jr{Phys. Rev. B} \textbf{85}(12), 125418 (2012).


\bibitem{ortiz_effective_2014}
 \textsc{G.~Ortiz},  \textsc{M.~Inchaussandague},  \textsc{D.~Skigin},
  \textsc{R.~Depine},  and  \textsc{W.\,L. Mochán},
 \jr{J. Opt.} \textbf{16}(10), 105012 (2014).


\bibitem{mendoza_tailored_2016}
 \textsc{B.\,S. Mendoza} and  \textsc{W.\,L. Mochán},
 \jr{Phys. Rev. B} \textbf{94}(19), 195137 (2016).


\bibitem{toranzos_optical_2017}
 \textsc{V.\,J. Toranzos},  \textsc{G.\,P. Ortiz},  \textsc{W.\,L. Mochán},
  and  \textsc{J.\,O. Zerbino},
 \jr{Mater. Res. Express} \textbf{4}(1), 015026 (2017).


\bibitem{perez-huerta_macroscopic_2013}
 \textsc{J.\,S. Pérez-Huerta},  \textsc{G.\,P. Ortiz},  \textsc{B.\,S.
  Mendoza},  and  \textsc{W.~Luis~Mochán},
 \jr{New Journal of Physics} \textbf{15}(4), 043037 (2013).


\bibitem{juarez-reyes_magnetic_2018}
 \textsc{L.~Juárez-Reyes} and  \textsc{W.\,L. Mochán},
 \jr{physica status solidi (b)} \textbf{255}(4), 1700495 (2018).


\bibitem{lapine_colloquium:_2014}
 \textsc{M.~Lapine},  \textsc{I.\,V. Shadrivov},  and  \textsc{Y.\,S.
  Kivshar},
 \jr{Rev. Mod. Phys.} \textbf{86}(3), 1093--1123 (2014).


\bibitem{czaplicki}
 \textsc{R.~Czaplicki},  \textsc{J.~Mäkitalo},  \textsc{R.~Siikanen},
  \textsc{H.~Husu},  \textsc{J.~Lehtolahti},  \textsc{M.~Kuittinen},  and
  \textsc{M.~Kauranen},
 \jr{Nano Letters} \textbf{15}(1), 530--534 (2015),
PMID: 25521745.


\bibitem{meza_second-harmonic_2019}
 \textsc{U.\,R. Meza},  \textsc{B.\,S. Mendoza},  and  \textsc{W.\,L.
  Mochán},
 \jr{Phys. Rev. B} \textbf{99}(12), 125408 (2019).


\bibitem{yu_flat_2014}
 \textsc{N.~Yu} and  \textsc{F.~Capasso},
 \jr{Nature Materials} \textbf{13}(2), 139--150 (2014).


\bibitem{yu_light_2011}
 \textsc{N.~Yu},  \textsc{P.~Genevet},  \textsc{M.\,A. Kats},
  \textsc{F.~Aieta},  \textsc{J.\,P. Tetienne},  \textsc{F.~Capasso},  and
  \textsc{Z.~Gaburro},
 \jr{Science} \textbf{334}(6054), 333--337 (2011).


\bibitem{khorasaninejad_metalenses_2016}
 \textsc{M.~Khorasaninejad},  \textsc{W.\,T. Chen},  \textsc{R.\,C. Devlin},
  \textsc{J.~Oh},  \textsc{A.\,Y. Zhu},  and  \textsc{F.~Capasso},
 \jr{Science} \textbf{352}(6290), 1190--1194 (2016).


\bibitem{chen_spin-controlled_2018}
 \textsc{Y.~Chen},  \textsc{X.~Yang},  and  \textsc{J.~Gao},
 \jr{Light Sci Appl} \textbf{7}(1), 1--10 (2018).


\bibitem{liren_deng_enhancing_2019}
 \textsc{{Liren Deng}},  \textsc{{Yanni Zhai}},  \textsc{{Yun Chen}},
  \textsc{{Ningning Wang}},  and  \textsc{{Yu Huang}},
 \jr{Journal of Physics D: Applied Physics} \textbf{52}(43), 43LT01 (2019).


\bibitem{paniagua-dominguez_ultra_2013}
 \textsc{R.~Paniagua-Domínguez},  \textsc{D.\,R. Abujetas},  and
  \textsc{J.\,A. Sánchez-Gil},
 \jr{Scientific Reports} \textbf{3}(March), 1507 (2013).


\bibitem{paniagua-dominguez_metallo-dielectric_2011}
 \textsc{R.~Paniagua-Domínguez},  \textsc{F.~López-Tejeira},
  \textsc{R.~Marqués},  and  \textsc{J.\,A. Sánchez-Gil},
 \jr{New J. Phys.} \textbf{13}(12), 123017 (2011).


\bibitem{abujetas_photonic_2015}
 \textsc{D.\,R. Abujetas},  \textsc{R.~Paniagua-Domínguez},
  \textsc{M.~Nieto-Vesperinas},  and  \textsc{J.\,A. Sánchez-Gil},
 \jr{J. Opt.} \textbf{17}(12), 125104 (2015).


\bibitem{ortiz_kellers_2018}
 \textsc{G.\,P. Ortiz} and  \textsc{W.\,L. Mochán},
 \jr{New J. Phys.} \textbf{20}(2), 023028 (2018).


\bibitem{keller_theorem_1964}
 \textsc{J.\,B. Keller},
 \jr{Journal of Mathematical Physics} \textbf{5}(4), 548--549 (1964).


\bibitem{s._mortola_two-dimensional_1985}
 \textsc{{S. Mortola}} and  \textsc{{S. Steffé}},
 \jr{Atti Accad. Naz. Lincei, Cl. Sci. Fis., Mat. Nat., Rend.} \textbf{78}, 77
  (1985).


\bibitem{milton_proof_2001}
 \textsc{G.\,W. Milton},
 \jr{Journal of Mathematical Physics} \textbf{42}(10), 4873--4882 (2001).


\bibitem{mochan_electromagnetic_1985}
 \textsc{W.\,L. Mochán} and  \textsc{R.\,G. Barrera},
 \jr{Phys. Rev. B} \textbf{32}(8), 4984--4988 (1985).


\bibitem{mochan_electromagnetic_II_1985}
 \textsc{W.\,L. Mochán} and  \textsc{R.\,G. Barrera},
 \jr{Phys. Rev. B} \textbf{32}(8), 4989--5001 (1985).


\bibitem{simon_analysis_1984}
 \textsc{H.\,D. Simon},
 \jr{Linear Algebra and its Applications} \textbf{61}(September), 101--131
  (1984).


\bibitem{horst_d._simon_lanczos_1984}
 \textsc{{Horst D. Simon}},
 \jr{Mathematics of Computation} \textbf{42}(165), 115--142 (1984).


\othercit
\bibitem{z._bai_lanczos_2000}
 \textsc{{Z. Bai}},  \textsc{{J. Demmel}},  \textsc{{J. Dongarra}},
  \textsc{{A. Ruhe}},  and  \textsc{{H. van der Vorst}},
Lanczos {Method},
 in: Templates for the {Solution} of {Algebraic} {Eigenvalue} {Problems}: {A}
  {Practical} {Guide},  (SIAM, Philadelphia, 2000).


\bibitem{glazebrook97pdl}
 \textsc{K.~Glazebrook} and  \textsc{F.~Economou},
 \jr{Dr. {Dobb's} Journal} \textbf{22}(9) (1997).


\othercit
\bibitem{moose}
 \textsc{S.~Ducasse},  \textsc{M.~Lanza},  and  \textsc{S.~Tichelaar},
Moose: An extensible language-independent environment for reengineering
  object-oriented systems,
 in: Proceedings of the Second International Symposium on Constructing Software
  Engineering Tools (CoSET 2000),  (IEEE, 2000).


\othercit
\bibitem{photonic}
 \textsc{W.\,L. Moch\'an},  \textsc{G.~Ortiz},  \textsc{B.\,S. Mendoza},  and
  \textsc{J.\,S. P\'erez-Huerta},
Photonic,
Comprehensive Perl Archive Network (CPAN), 2016,
Perl package for calculations on metamaterials and photonic structures.


\end{thebibliography}
%



\newpage

\section*{Graphical Table of Contents\\}
GTOC image:
\begin{figure}[h]%
\includegraphics[width=4cm,height=4cm]{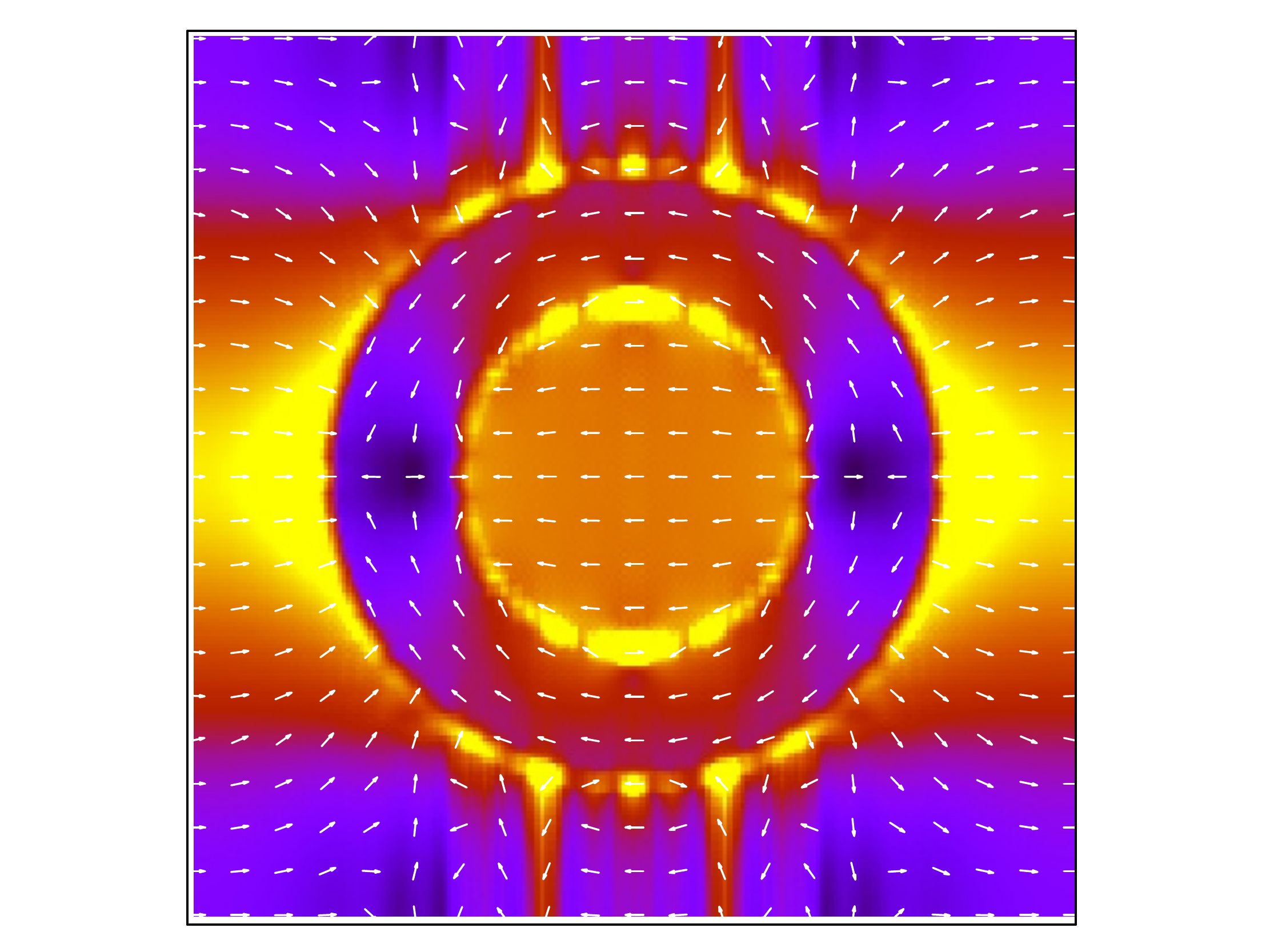}
\caption*{%
  We develop an efficient procedure to compute recursively the fields and the
  response of metamaterials made of any number of components with an
  arbitrary geometry and composition, metallic or dielectric,
  transparent or dissipative, using a spinor-like representation of
  the field states. We test the procedure against systems with
  analytically determined properties. The figure shows the field of a
  square array of Ag covered SiO$_2$ cylinders close to resonance.
}
\label{GTOC}
\end{figure}

\end{document}